# Nonlinear laser lithography as a new technology for the high-quality alignment of nematic liquid crystals

I A Pavlov[a,b], A S Rybak[b], A M Dobrovolskiy[b], V M Kadan[b], I V Blonskiy[b], F Ö Ilday[a,c], Z I Kazantseva[d], I A Gvozdovskyy[b]

[a]Bilkent University, 06800 Çankaya, Ankara, Turkey

[b]Institute of Physics, NAS of Ukraine, Prospekt Nauki 46, Kyiv-28, 03028, Ukraine

[c]UNAM - National Nanotechnology Research Center and Institute of Materials Science and Nanotechnology, Bilkent University, 06800 Çankaya, Ankara, Turkey

[d]V. E. Lashkaryov Institute of Semiconductor Physics, NAS of Ukraine, Prospekt Nauki 41, Kyiv-28, 03028, Ukraine

E-mails:

pavlov.iop@gmail.com (I. A. Pavlov); rybak.andrey@gmail.com (A. S. Rybak); dobr@iop.kiev.ua (A. M. Dobrovolskiy); kadan@iop.kiev.ua (V. M. Kadan); blon@iop.kiev.ua (I. V. Blonskiy); ilday@bilkent.edu.tr (F. Ö. Ilday); kazants@isp.kiev.ua (Z. I. Kazantseva); igvozd@gmail.com (I. A. Gvozdovskyy)

**ABSTRACT**

It is well known that today two main and well studied methods for alignment of liquid crystals has been used, namely: rubbing and photoalignment technologies, that lead to the change of anisotropic properties of aligning layers and long-range interaction of the liquid crystal molecules in a mesophase. In this manuscript, we propose the usage of the nonlinear laser lithography technique, which was recently presented as a fast, relatively low-cost method for a large area micro- and nanogrooves fabrication based on laser-induced periodic surface structuring, as a new perspective method of the alignment of nematic liquid crystals. 920 nm periodic grooves were formed on a Ti layer processed by means of the nonlinear laser lithography and studied as an aligning layer. Aligning properties of the periodic structures of Ti layers were examined by using a combined twist LC cell. In addition, the layer of the microstructured Ti was coated with an oxidianiline-polyimide film with annealing of the polymer film followed without any further processing. The dependence of the twist angle of LC cells on a scanning speed and power of laser beam during processing of the Ti layer was studied. The azimuthal anchoring energy of Ti layers with a periodic microstructure was calculated. The maximum azimuthal anchoring energy for the microstructured Ti layer was about $4.6 \times 10^{-6}$ J/m$^2$, which is comparable to the photoalignment technology. It was found that after the deposition of a polyimide film on the periodic microstructured Ti layer, the gain effect of the azimuthal anchoring energy to $\sim 1 \times 10^{-4}$ J/m$^2$ is observed. Also, AFM study of aligning surfaces was carried out.

Corresponding author.

E-mail address: igvozd@gmail.com (I.A.Gvozdovskyy).

# 1 Introduction

The alignment of liquid crystals (LCs) is important and a key condition for their application as a liquid with anisotropic properties in manufacturing LC displays (LCD) and different devices. Due to, on the one hand, a long range of orientational interaction and relatively free movement of anisotropic LC molecules in mesophase and, on the other hand, the creation of anisotropic properties of aligning layers, the homogeneous alignment of LC bulk on the macroscopic scale has been observed. The creation, study and characterization of the alignment surface, obtained by means of different methods, are an important task to the LC practical application. For this purpose, both the different aligning materials and various methods of their processing can be used, as demonstrated in many references in book [1] and reviews [2-5]. However, in present there are two main methods, well studied and widely used to create the aligning layers for further application in LCD technology, *etc*.

The first method, used extensively for different applications in industry, is the rubbing technique with various materials [1-9] of different surfaces [1,10]. However, in spite of the fact that the rubbing technology is widely used in LCD technology, this technique has some shortcomings, among which are accumulation of both the static charges and dust particles [4].

The latter method is a so-called photoalignment effect, which was for the first time described by K. Ishimura [10] for azobenzene layers controlling the LC alignment with light in zenithal plane. The homogeneous aligning of LCs in an azimuthal plane of aligning films of the substrate was simultaneously discovered by groups of W. Gibbons [12], M. Schadt [13] and Yu. Reznikov [14,15]. As was shown, the photoaligning technique is a really alternative method to the rubbing technique, because the usage of photosensitive materials, deposited on a substrate or dissolved in bulk of LCs [16], leads to the change of the orientational order of photoproducts under polarized light irradiation. In addition, the usage of a plasma beam as a perspective method of processing aligning layers for the homogeneous planar and tilted orientation of LCs was recently studied in [17,18]. In the case of the photoaligning of LCs, the mechanical contact with the surface of a substrate, with all the shortcomings that this entails [4], is absent. As noted in review [4], the photoaligning technology gives an effective control of main anchoring parameters (easy orientation axis, pretilt angle and anchoring energy). In addition, the value of the anchoring energy of the photoaligning surfaces strongly depends on chemical properties of the using materials and can be within the wide range $10^{-8} - 10^{-5}$ J/m$^2$ [4].

In the case of above-mentioned methods, the formation of the aligning layers can be achieved by means of deposition of inorganic materials, by polymer-coating from different solutions (Langmuir-Blodgett, spin-coating or dipping technique) followed by the high-temperature process (for instance, polymerization) and further using the rubbing roll or the polarized light. As shown in [5-7,18,19] after a certain method of the processing of the aligning layers, the period of the ripple structure can change within 100 – 300 nm, while the amplitude (depth) of relief can be within about 80 - 150 nm.

In addition, to obtain the LC alignment by means of surface with a small period (about 235 -250 nm) of nano-grooves, the e-beam lithography [20] and AFM nano-rubbing [21] were applied. However, the area of the nano-grooves rubbed on the surface was very small (about 15 μm long and 400 μm wide) [20], and both methods show a very low throughput. Moreover, nano-imprint lithography [22] and photolithography [23] were also used to create nano-grooves on a polymer surface for aligning LC molecules, but both techniques are complicated for the preparation of masks, and the period of nano-grooves is limited.

It should be also noted that recently, a fast and high-throughput method, consisting of the splitting of a polymer film with further propagation of the wave front to induce self-assembled micro and nanogrooves on a polymer surface (the so-called crack-induced grooving method or CIG method), was proposed to align LCs [24]. This method does not require additional high-temperature processing (about 250° C) of aligning layers. The usage of this method avoids the presence of dust and surface charge (many ions) on the surface of aligning layers. CIG method also provides a relatively large anchoring energy within a range $10^{-6} - 10^{-5}$ J/m$^2$ [24], comparable to that having various polymers both after the rubbing and photoaligning process [4].

In this manuscript, we have made aligning layers, using a simple, high-speed and low-cost method for a large area micro and nanogrooves fabrication based on a laser-induced periodic surface structuring with femtosecond pulse, which is well known as nonlinear laser lithography (NLL) [25]. As a model of aligning layer the metal-oxide periodic microstructures (or so-called ripples) were examined. Contrary to the traditional methods [1-19] of the LC alignment, the preparation of the aligning surface can consist mostly of two stages. At the first main stage, the processing of the deposition of a titanium (Ti) layer on a glass substrate by NLL method results in the creation of the periodic microgrooves with certain parameters (for instance, depth, period and angle of direction of grooves). At the second additional stage of the creation of aligning surfaces, the Ti layer was coated with a polymer followed by the polymerization process without any processing (such as rubbing or irradiation with polarized light). Here, the homogeneous alignment of the nematic liquid crystal and the dependence of the anchoring energy in the azimuthal plane of aligning films on various parameters of NLL method for both the pure microstructured Ti layer and similar layer with the polymer-coated surface are described.

## 2 Materials and methods

### 2.1 Materials

To study the aligning properties of the structured Ti layers, the nematic liquid crystal E7, obtained by Licrystal, Merck (Darmstadt, Germany) was chosen. The optical and dielectrical anisotropy of the nematic E7 at T = 20 °C, λ = 589.3 nm, and $f$ = 1 kHz are $\Delta n$ = 0.2255 ($n_e$ = 1.7472, $n_o$ = 1.5217) and $\Delta\varepsilon$ = +13.8, respectively. Splay, twist and bend elastic constants of nematic E7 are $K_{11}$ = 11.7 pN, $K_{22}$ = 6.8 pN, $K_{33}$ = 17.8 pN, respectively [26-28].

To obtain the planar alignment in the azimuthal plane of the nematic liquid crystal E7, both the polyimide PI2555 (HD MicroSystems, USA) and 1-% DMF solution of oxidianiline-polyimide (ODAPI) (Kapton synthesized by I. Gerus, Institute of Bio-organic Chemistry and Petrochemistry, NAS of Ukraine) were used.

### 2.2 Methods

#### 2.2.1 Preparation of microstructured Ti layer

To examine the aligning properties of the metal-oxide microstructure, Ti layers, recently studied in [25], were chosen as a model. For this aim, a 300 nm thick Ti layer was deposited on a glass substrate. To create the large area of structured Ti layers we used the experimental scheme of the NLL method, described in [29]. As schematically shown in Figure 1 (a), the setup consists of a home-made femtosecond fiber laser system, described in [29], galvanometer-scanner and motorized 3D-translation stage. The laser can produce up to 1 μJ of pulse energy at repetition rate of 1 MHz which corresponds to 1 W of the average power. The minimal pulse duration which can be obtained from the system, is 100 fs, however we found that the increase of the pulse duration up to several hundred femtoseconds does not have any observable effect on the structure formation. The half-wave plate (HWP) placed before the polarization beam splitter (PBS) provides control of the laser power on a sample. The second HWP allows control of polarization on a sample. The sample was placed on motorized 3D stage in the focal plane of galvanometer-scanner`s f-theta lens.

The beam was raster-scanned over the sample surface as shown in Figure 1 (b). The laser spot is schematically shown as a solid "pink" circle. The polarization direction is shown on the picture as a direction of E vector. The pulse energy, scanning speed and the overlap factor were adjusted to preserve the concept of NLL, that allows coherent extension of the structure over a large area surface. In this case, the small beam with a 9 μm diameter preserves coherency of the scattered light within the beam spot, and the scanning introduces a positive nonlocal feedback from the already created structure to the new area on the surface. The maximum scanning area for our model of the galvanometer-scanner is 1×1 cm$^2$. With the pulse energy equal 0.35 μJ and 1 MHz repetition rate of the laser, the processing speed was as fast as 7 second for 5×5 mm$^2$ area. The macro-photograph of the sample with ten structured areas and SEM image of the structured Ti layer, having square area with dimension 5×5 mm$^2$, are shown in Figure 1 (c).

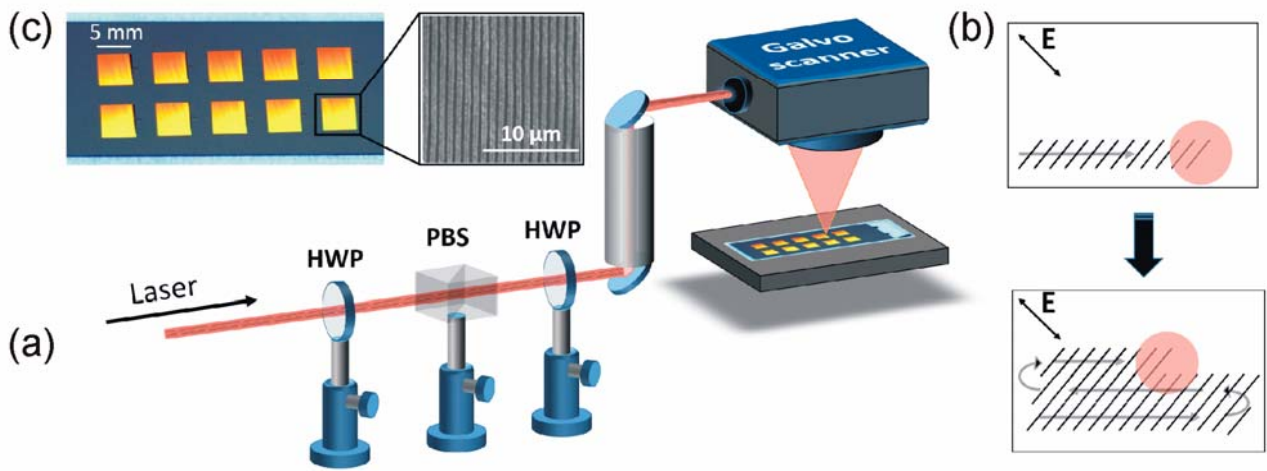

Fig. 1. (a) Scheme of the structuring of the Ti layer by NLL method. (b) Cartoon, demonstrating scanning direction of the laser beam over the sample during NLL process. (c) Photograph of the sample with ten structured areas, and SEM image of the square area with dimension 5×5 mm$^2$.

NLL method allows us to do changes in a wide range of the scanning speed $v$ = 500 – 3000 mm/s and laser power $P$ within a range 150 – 375 mW. An additional point to emphasize is that at the power level less than 150 mW the structure is not created, while at the power level is more than 375 mW the structure is generally ruined. After processing of the Ti layer by NLL method, the direction of microstructures $θ_1$ was inclined at an angle of 8° to the horizontal side of the square area, as schematically shown in Figure 2.

It is well known that the changes in wavelength of the laser may change the period of microgrooves [30,31]. However, in our studies the period of microgrooves was constant. The value of the period was ~ 920 nm obtained by means of the direct AFM measurements.

*2.2.2 Preparation of aligning films*

To estimate the value of the azimuthal anchoring energy of microstructured Ti layers we used the idea of the combined twist LC cell [32-34], consisting of the tested and reference substrates. For this aim the polyimide PI2555, possessing a strong anchoring energy [35,36], was used for the preparation of reference substrates, while the ODAPI was utilized for the preparation of tested plates. The polyimide PI2555 film was deposited on glass substrates by the spin-coating method (6000 rpm, 30 s). About 50 nm ODAPI films at the microstructured Ti layers have been formed by means of dipping technique using equipment for Langmuir-Blodgett film preparation R&K (Wiesbaden, Germany). For this, tested substrates were dipped into an appropriable solution and further vertically drown up at the constant speed about 5 mm/min along or across the direction of the microstructured Ti layers. The polyimide PI2555 film (reference plate) was annealed at 180° C for 30 min. The microstructured Ti layer with deposited ODAPI film (tested plate) was annealed at 190° C for 90 min. Thereafter, only polyimide PI2555 films were unidirectionally rubbed ($N_{rubb}$ = 10 times) with the pressure of rubbing 850 N/m$^2$, to reach a strong anchoring energy $W \sim (4 \pm 1) \times 10^{-4}$ J/m$^2$ [35,36]. The direction of rubbing PI2555 films of the reference substrates made an angle $\theta_2 = 45°$ with the horizontal side of the square area (microstructured Ti layer) of the tested plates, as shown in Figure 2.

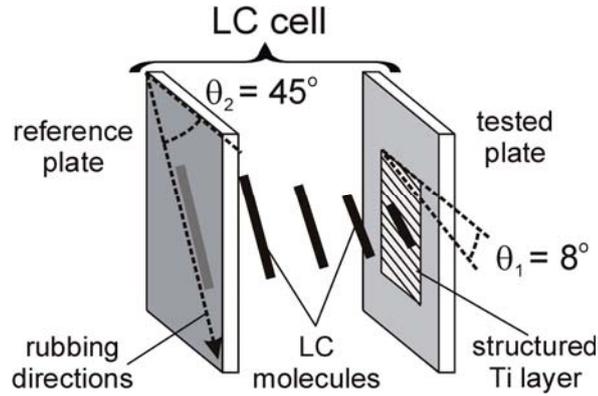

Fig. 2. Schematic image of the combined twist LC cell, consisting of the reference substrate (rubbed PI2555 film) and tested substrate (microstructured pure Ti layer or coated with ODAPI film). Direction of the microgrooves structured Ti layer $\theta_1$ is inclined at 8° to the horizontal side of the square area. Direction of rubbing of the reference substrate $\theta_2$ is at 45° with the horizontal side of the square area (microstructured Ti layer) of the tested plate.

*2.2.3 Preparation of the combined twist LC cell*

As was proposed in [32-34], to measure the twist angle and further calculate the azimuthal anchoring energy of the studying microstructured Ti layers we made combined twist LC cells. LC cells consisted of the tested and reference substrates as can been see in Figure 2. The tested substrates were used of two types. The first type of the tested substrate was coated with a Ti layer and further processed by the NLL method. The second type of the tested substrate consists of the first type substrate additionally coated with a 1-% DMF solution of ODAPI. The reference substrate was coated with a polyimide PI2555 processed with the rubbing technique.

The easy axis of the two tested and reference substrates is given by the direction of rubbing on the one hand, at a 45° angle to the horizontal side of the reference substrate and on the other hand, along the microgrooves of the tested substrate (Fig. 2). In this case the angle $\varphi_0$ between the easy axis of the reference and the tested substrate is $\theta_2 - \theta_1 = 36°$.

The thickness of a gap was set to 20 – 25 μm by a Mylar spacer and measured by means of the interference method, using transmission spectra of empty LC cells. The LC cells were filled with the nematic LC at the higher temperature T = 61° C than the temperature of the isotropic phase ($T_{Iso}$ = 58° C) [26] and slowly (~ 0.1° C/min) cooled to the room temperature to avoid the possible flow alignment.

*2.2.4 Measurement of the twist angle and calculation of the azimuthal anchoring energy*

An experimental setup for the measurement of the twist angle $\varphi$ of the combined twist LC cells is shown in Figure 3 (a). For this aim the linear-polarized light (TEM$_{00}$, $\lambda$ = 632.8 nm and power $P$ = 1.5 mW) from He-Ne laser LGN-207a (Lviv, Ukraine) and a silicon photodetector (PD) FD-18K (Kyiv, Ukraine) with the spectral range 470 - 1100 nm were used. PD was connected to the oscilloscope Hewlett Packard 54602B 150MHz (USA).

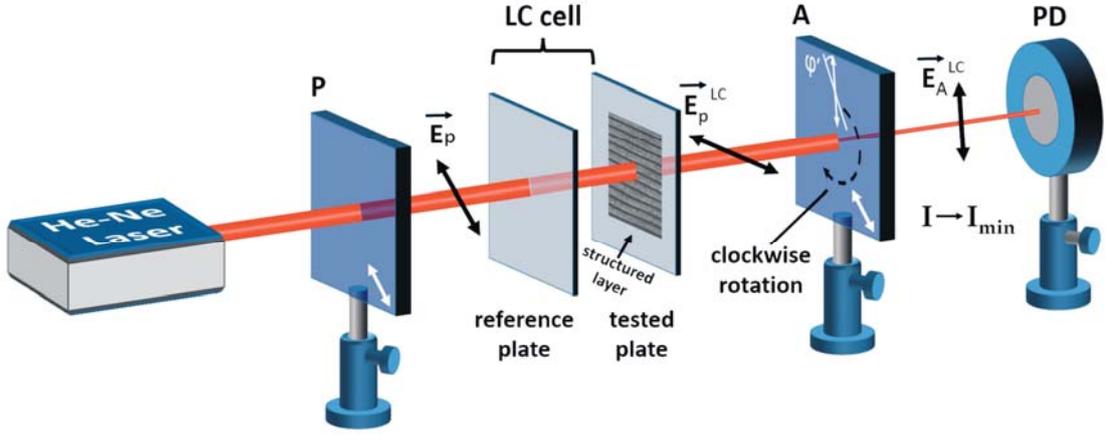

Fig. 3. Scheme of the measurement of the twist angle $\varphi$ of the combined twist LC cell, assembled with the reference PI2555 ($N_{rubb}$ = 10) and tested substrates of both types.

The twist LC cell (Fig. 2) was placed between a pair of parallel polarizes (*i.e.* the polarization plane of a polarizer (P) and analyzer (A) coincides), as shown in Figure 3. The rubbing direction of the reference substrate (PI2555 film) coincides with the polarization plane of a pair of polarizers. The direction of microstructured Ti layers for both types of tested substrates was at an angle $\varphi_0 = 36°$ to the polarization plane of P and A. The linear-polarized incident beam passes from the He-Ne laser through P and the LC cell toward A and further to the photodetector PD. In the twist LC cell the laser beam $\vec{E}_P$ is rotated at a certain twist angle $\varphi$, which depends on the azimuthal anchoring energy $W$ of the tested substrate [32,33]. Behind the LC cell the rotated beam $\vec{E}_P^{LC}$ propagates toward the analyzer (A), as can be seen in Figure 3 (a). Due to the different orientation of the polarization plane between the rotated beam $\vec{E}_P^{LC}$ (after it passed through the twist LC cell) and the polarization plane of the analyzer (A), a certain part $\vec{E}_A^{LC}$ of the beam $\vec{E}_P^{LC}$ will reach the photodetector PD. In order to measure a value of the twist angle $\varphi$, depending on the azimuthal anchoring energy $W$ of the tested substrate [34], the analyzer (A) was clockwise rotated through some angle $\varphi'$ to obtain the minimal intensity (*i.e.* $I \to I_{min}$) of the incident beam on PD connected to the oscilloscope. In this case the value of the angle $\varphi'$ is the real twist angle $\varphi$ between the easy axis of the reference and tested substrates.

Measurements of the twist angle $\varphi$ of the sample allows us to calculate the value of the azimuthal anchoring energy ($W_\varphi$) of the aligning layer of the tested substrates. According to [32,34] the twist angle $\varphi$ is related to the azimuthal anchoring energy $W_\varphi$ as follows:

$$W_\varphi = K_{22} \frac{2\sin(\varphi)}{d \sin 2(\varphi_0 - \varphi)}, \quad (1)$$

where $d$ is the thickness of the LC cell, $\varphi_0 = 36^\circ$ is the angle between the easy axes of the reference and tested substrates and $\varphi$ is the measured twist angle.

**3 Results and discussion**

A typical AFM image of the structured Ti layers after processing with the help of the NLL method is shown in Figure 4 (a). It is obvious that the usage of the micro-periodic structures obtained by the NLL method after processing of the various metal films can be applied to the creation of aligning layers.

As can be seen from Figure 4 (a) the certain periodic structure of grooves (ripples), characterized by a certain period ($\Lambda$) and depth of grooves (A), is formed. By using AFM studies it was found that the depth of grooves depends on main parameters (speed of scanning $\upsilon$ and laser power P), which are set before the processing of the Ti layer by the NLL method.

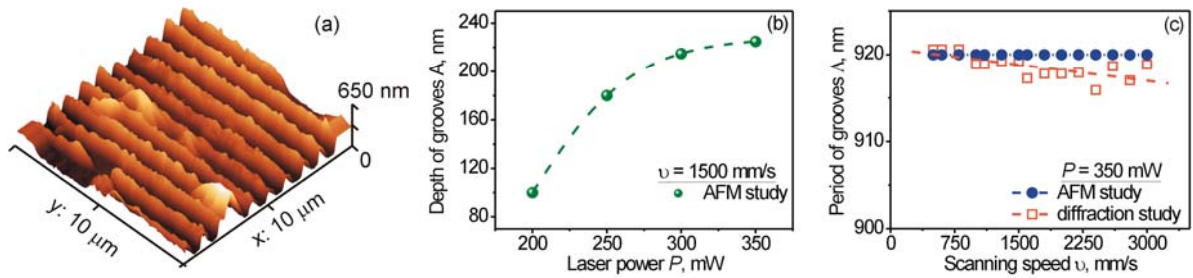

Fig. 4. (a) AFM image of the structured Ti layer after processing by the NLL method with a speed of scanning $\upsilon = 1500$ mm/s and laser power $P = 350$ mW. (b) Dependence of the depth of grooves A on laser power $P$ at a constant scanning speed $\upsilon = 1500$ mm/s. (c) Dependence of the period of grooves $\Lambda$ on a scanning speed at a constant laser power $P = 350$ mW, measured by the AFM method (solid circles) and diffraction method (open squares). The dashed line is a guide to the eye.

The dependence of the depth of grooves A on a laser power $P$ at a constant scanning speed $\upsilon = 1500$ mm/s is shown in Figure 4 (b). It is seen that the increase in the laser power of the NLL method leads to a rise of the depth of grooves.

Since the microstructured Ti layer is a diffraction grating than according to [36] for the normal incident beam, the grating period $\Lambda$ is as follows:

$$\Lambda = m \cdot \lambda / \sin \alpha , \qquad (2)$$

where $m$ is the diffraction order, $\lambda$ and $\alpha$ is the wavelength and the angle of diffraction of the incident beam, respectively.

By means of both the AFM and diffraction studies, it was also found that the change of the laser power or the speed of scanning does not strongly influence the change of the value of the period of grooves $\Lambda$ (Fig. 4(c)). Although the values of the period of microgrooves measured by both methods are a little different (Fig. 4 (c)), however, the accuracy of the AFM method is higher, owing to the direct measurements contrary to the diffraction method (Eq. 2). It will be recalled that the microstructured period can be only changed by the usage of a laser with a different wavelength [30,31] or by a tilt angle, but it is difficult for our setup.

In contrast to the traditionally used methods of the LC alignment [6,7,18,19], where the period of grooves changes within a range 100 – 300 nm, we think that despite the fact that processed Ti layers, having the period of grooves about 920 nm, may produce the homogeneous alignment of nematic, however, their azimuthal anchoring energy may be sufficiently low.

For this aim, to estimate the value of the anchoring energy of the structured Ti layer we use the well-known Berreman's theory [6,7]. This theory supposes that the anchoring energy depends on the depth A and period Λ of grooves as follows [6,7,18,24]:

$$W_B = 2\pi^3 \cdot A^2 \cdot K_{22} / \Lambda^3 \qquad (3)$$

According to Equation (3), dependencies of the anchoring energy $W_B$ on the change of both the period (Λ) and depth (A) of grooves, are shown in Figure 5 (a) and Figure 5 (b), respectively.

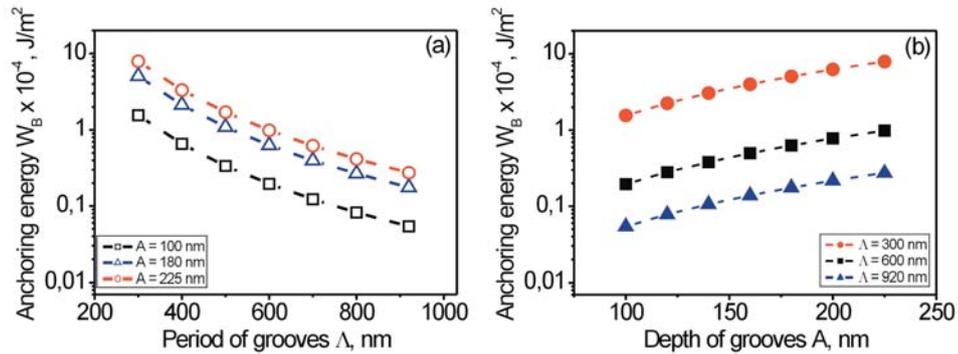

Fig. 5. Dependence of the anchoring energy, calculated by Berreman's theory, of the aligning layer on: (a) period (open symbols) and (b) depth of grooves (solid symbols). The dashed line is a guide to the eye.

As can been see from Figure 5, the maximum value of the anchoring energy can be obtained for aligning layers, having on the one hand, a small value of the period of grooves and on the other hand, large value of the depth of grooves.

Let us estimate the maximum value of the anchoring energy of the microstructured Ti layer, by using the Equation (3) and values of the period and depth of grooves measured by AFM (Fig. 4 (b) and Fig. 4 (c)). The maximum value $W_B$ of the microstructured Ti layer (Λ ~ 920 nm and A ~ 225 nm) at certain parameters of NLL processing (laser power $P$ = 350mW and speed of scanning $v$ = 1500 mm/s) reaches ~ $2.7 \times 10^{-5}$ J/m$^2$. It should be noted that contrary to [41], for the tested substrate of the first type, the depth of grooves A is higher due to the usage of a thicker Ti layer (~ 300 nm). The obtained value $W_B$ is of order of the azimuthal anchoring energy of azopolymers [27], photo-crosslinking materials [38,39] and polyvinylcinnamate (PVCN) [16,40], and it is in a good agreement with the anchoring energy for Λ ~ 800 nm and A ~ 100 nm received by the crack-induced grooving (CIG) method [24]. The calculated dependencies of the anchoring energy $W_B$ (Eq. 3) of the microstructured Ti layer on both the power of laser beam $P$ and depth of grooves A, at a constant period of grooves Λ = 920 nm, are shown in Figure 6 (a) and Figure 6 (b), respectively.

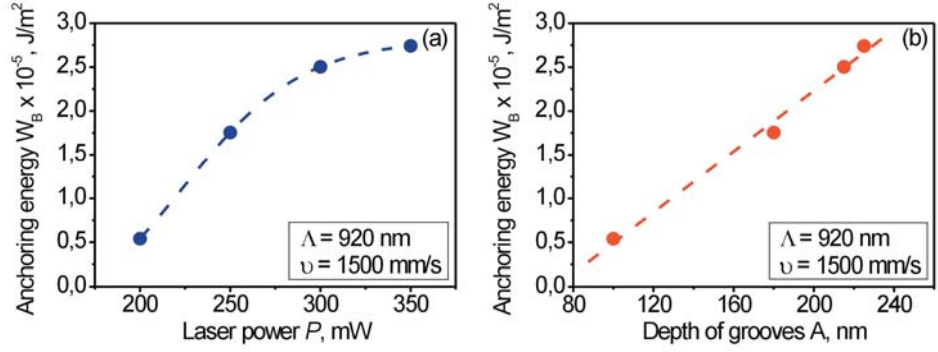

Fig. 6. Dependencies of the anchoring energy $W_B$ of the microstructured Ti layer on: (a) the power of laser beam $P$ and (b) depth of grooves A. The period of grooves of the microstructured Ti layer is 920 nm, and the scanning speed is 1500 mm/s. The dashed line is a guide to the eye.

As can be seen from Figure 6 (a), that the increase of the laser power $P$ in the NLL method results in the increase of the anchoring energy $W_B$ (Eq. 3), owing to the increase of the grooves depth A (Fig. 4 (b)). With the strong laser power $P$ within a range ~ 280 – 350 mW, both the depth of grooves A (Fig. 4 (b)) and anchoring energy $W_B$ (Fig. 6 (a)) reach their maximum value. At certain conditions (*i.e.* wavelength of beam, scanning speed) of the NLL processing of the Ti layer, hypothetically the further growth of the anchoring energy $W_B$ can be reached by the increase of the depth of grooves more than at ~ 230 nm, as it is easy to see from Figure 6 (b).

It is obvious that the value of the anchoring energy $W_B$, at constant parameters of the NLL structuring (under the experimental conditions) and without the change of the laser wavelength (e.g., the usage of a second harmonic generation), that leads to the decrease of the period of microgrooves, can not be changed. However, to increase the anchoring energy the usage of a microstructured Ti layer additionally coated with an ODAPI film was recently proposed in [41]. Contrary to [41], here we obtained the tested substrate of the second type with the depth of grooves within a range A ~ 150 - 200 nm (for further estimations A = 175 nm) by a trial-and-error method in the dipping technique, but the average value of the period $\Lambda$ ~ 900 nm is identical with its in [41]. For these tested substrates the greatest possible calculated value of the anchoring energy $W_B$ is within a range ~ $(1.3 – 2.3) \times 10^{-5}$ J/m$^2$, owing to both the insignificant changes in the period and certain decrease of the depth of grooves after it has been additionally coated with an ODAPI film. It is obvious that the anchoring energy does not only depend on the period and depth of grooves, but also depends on physical and chemical properties of the surface [25,37-39].

Now let us consider the quality of the alignment of nematic E7 by structured microgrooves, using two types of the tested substrates and also calculate the real value of the azimuthal anchoring energy $W_\varphi$ from the experimentally measured twist angle, by using the method of the combined twist LC cell [32-34].

Figure 7 shows photographs of two different combined twist LC cells, consisting of, on the one hand, the reference substrate and, on other hand, of two types of the tested substrates. LC cells placed between a pair of polarizers with different angles of planes of polarization. As can be seen from photographs, the homogeneous alignment of the nematic LC is observed for the twist LC cell, consisting of tested substrates, having both the microstructured Ti layer (Fig. 7 (a-c)) and the microstructured Ti layer coated with the ODAPI film (Fig. 7 (d-f)). In this case the homogeneity alignment was achieved by, on the one hand, microgrooves of both the Ti layers and Ti layers coated with the ODAPI film and, on the other hand, the rubbed surface of PI2555. We have considered the microstructures obtained by the NLL method, as an analogue to the rubbed (or photoaligning) surface with a difference that the period of microgrooves for the rubbing (or photoalignment) technique is far less than the usage of the NLL [41] or CIG method in [24].

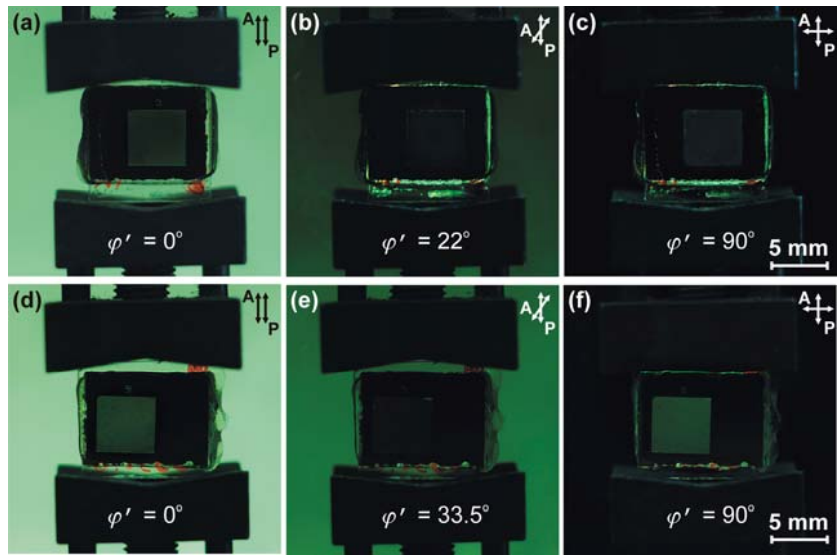

Fig. 7. Photographs of the twist LC cells placed between a pair of polarizers with different angles of planes of polarization: (a), (d) - parallel; (b), (e) – twist angle $\varphi'$; (c), (f) - perpendicular. The LC cells consist of the reference (rubbed PI2555 film) and tested substrate, having the microstructured Ti layer (a-d) and microstructured Ti layer coated with the ODAPI film (d-f). The twist angle $\varphi'$ for the LC cell: (a-c) - 22°; (d-f) – 33.5°. The thickness of the twisted LC cell was: (a) – (c) $d = 23.3$ μm; (d) – (f) $d = 23.8$ μm. The processing of Ti layers was carried out by the NLL method at the scanning speed $\upsilon = 500$ mm/s and laser power $P = 250$ mW.

To check our statement that the homogeneous alignment is observed owing to on the one hand, the availability of microgrooves and on the other hand, the usage of the ODAPI film deposited on microgrooves, we made two combined twist LC cell, consisting of the reference substrate (rubbed or non-rubbed PI2555) and tested substrate (non-rubbed ODAPI or ODAPI deposited on the microstructured Ti layer). As can be seen from Figure 8 a low quality of the alignment of nematic E7 for these twist LC cells is observed.

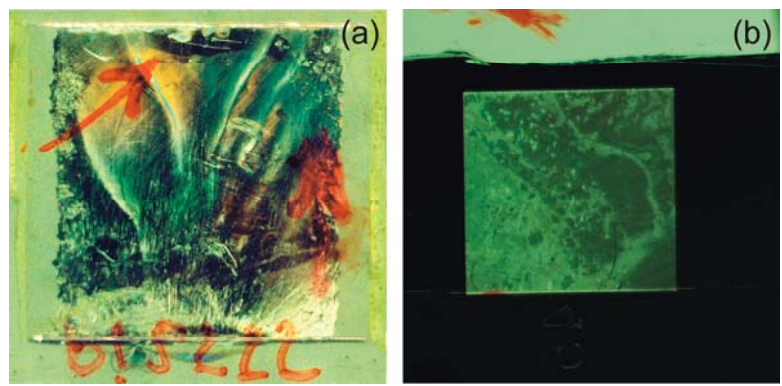

Fig. 8. Photographs of the twist LC cells placed between parallel polarizers. LC cells consist with a pair of substrates: (a) the reference glass plate is the rubbed PI2555 film and tested plate is the non-rubbed ODAPI film formed by the dipping technique; (b) the reference plate is the non-rubbed PI2555 film and tested plate is the microstructured Ti layer coated with the ODAPI film by means of the dipping technique. Thickness of LC cells was ~ 24.2 μm.

To calculate the value of the azimuthal anchoring energy of combined twist LC cells, at the beginning the measurements of twist angles were carried out. Since the NLL method [29] allows to do changes in a wide range of the scanning speed $v = 500 – 3000$ mm/s and laser power $P$ within a range $150 – 375$ mW, then we studied dependencies of twist angles $\varphi(v)$ and $\varphi(P)$ (Fig. 9), which can be analogue to, for instance, dependencies of the twist angle on a number of times of unidirectional rubbings $N_{rubb}$ [34] and the pressure of rubbing $p_{rubb}$ [35,36] in the rubbing technique of alignment of the nematic LCs.

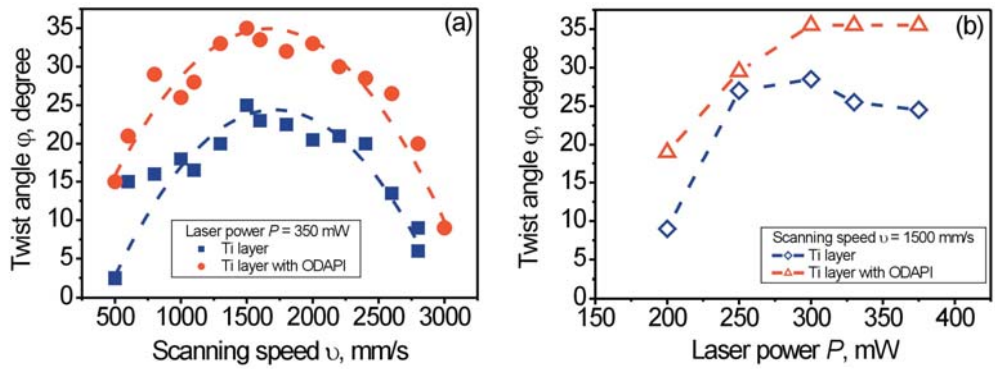

Fig. 9. Dependence of the twist angle $\varphi$ of the LC cell on: (a) the scanning speed $v$ and (b) laser power $P$. The LC cell consists of the both types of the tested substrate: microstructured Ti layer (solid squares and open diamonds) and microstructured Ti layer coated with the ODAPI film (solid circles and open triangles). The dashed line is a guide to the eye.

As can been see from Figure 9 (a) the increase of the scanning speed in the NLL technique at a constant laser power ($P = 350$ mW) leads to the non-monotonically changes of the twist angle for both types of the tested substrates [41]. The increase of the laser power $P$ leads to the monotonous increase of the twist angle of the LC cell at a constant scanning speed in the NLL method, as is shown in Figure 9 (b). It is seen that for both types of the tested substrates, twist angles of the LC cells reach maximum values at a certain optimal range of values of $v = 1300 – 2000$ mm/s and $P = 250 – 375$ mW. In addition, it should be noted that the usage of the ODAPI film (tested substrates of the second type) leads to the growth of the twist angle, as shown in Figure 9 (a) and Figure 9 (b), depicted by solid circles and open triangles, respectively.

By knowing the value of the twist angle $\varphi$, we calculated the azimuthal anchoring energy $W_\varphi$, by using Equation 1. For both types of the tested substrates the dependence of the azimuthal anchoring energy $W_\varphi$ on the value of the scanning speed $v$ is shown in Figure 10.

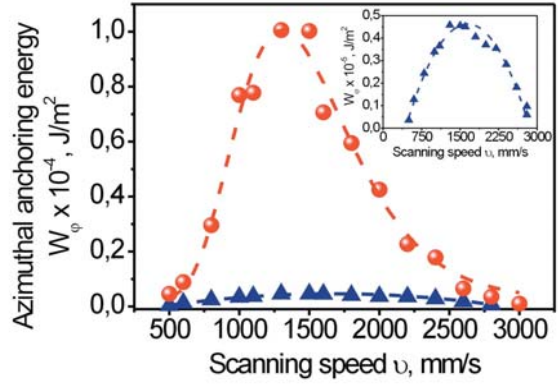

Fig. 10. Dependence of the calculated azimuthal anchoring energy $W_\varphi$ on the scanning speed $v$ for LC cells consisting of the both types of the tested substrates, having the microstructured Ti layer (solid triangles) and microstructured Ti layer coated with the ODAPI film (solid spheres). The inset depicts the dependence of $W_\varphi(v)$ for the microstructured Ti layer in a large scale. The dashed line is a guide to the eye.

It is seen that the azimuthal anchoring energy $W_\varphi$ of the tested substrate with the microstructured Ti layer reaches the value $\sim 0.46\times10^{-5}$ J/m$^2$ (see also inset), which is $\sim 6$ times low than the value obtained by Berreman's theory [6,7]. It may be safely suggested that the difference between values of anchoring energies, obtained with Equation (1) and Berreman's theory (Eq. 3), is observed due to the fact, that the theory does not take into account the interaction between LC molecules and the aligning surface. However, as was mentioned above, the value of the azimuthal anchoring energy $W_\varphi$ corresponds to the azimuthal anchoring energy of photoaligning layers [39-41].

In the case of the structured Ti layer coated with the ODAPI film, the dependence of the azimuthal anchoring energy on the scanning speed $v$ is shown in Figure 10 with solid "red" spheres. It is seen that coating of the polymer film onto the structured Ti layer leads to the dramatic increase in the azimuthal anchoring energy value $\sim 1\times10^{-4}$ J/m$^2$. This value is approximately 22 times greater than the anchoring energy of the pure microstructured Ti layer. A strong value increase of the azimuthal anchoring energy of the microstructured Ti layer coated with the OPADI film can be explained, on the one hand, by availability of the depth of grooves (as was described above, the average value of energy is $\sim 1.8\times10^{-5}$ J/m$^2$ for instance at $\Lambda = 900$ nm and $A = 175$ nm) and, on the other hand, owing to the usage of the polymer which enhances the interaction between molecules of the liquid crystal and polymer (this contribution is 4.6 times greater than for the microstructured Ti layer and is $\sim 8.2\times10^{-5}$ J/m$^2$).

As can be seen from Figure 9 (b), the change of the laser power $P$, used to the processing of the Ti layer at a certain constant scanning speed $v$ in the NLL method, has a strong impact on the value of the twist angle and thus on the value of the azimuthal anchoring energy, as shown in Figure 11.

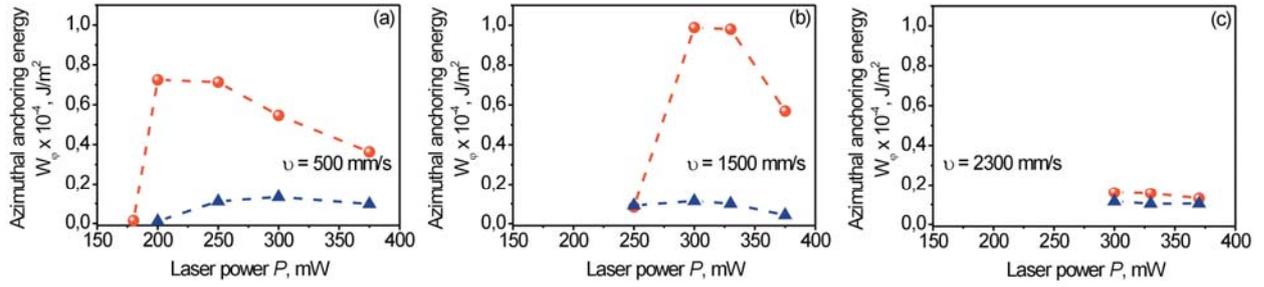

Fig. 11. Dependence of the calculated azimuthal AE (Eq.1) of aligning films on the power of the laser beam $P$ at a constant scanning speed $\upsilon$: (a) 500 mm/s, (b) 1500 mm/s and (c) 2300 mm/s during the processing of the Ti layer with the NLL method. The LC cell consists of the tested substrate having the microstructured Ti layer (solid triangles) and structured Ti layer coated with the ODAPI film (solid spheres). The dashed line is a guide to the eye.

As it is easy to see, there is an optimal value of the laser power $P$ for each constant scanning speed $\upsilon$, when the highest possible value of the azimuthal anchoring energy reaches the order of $\sim 10^{-5}$ J/m$^2$ (Fig. 11, solid triangles). As was mentioned above, to enhance the azimuthal anchoring energy of microstructured Ti layers, they additionally were coated with the ODAPI film with the process of polymerization followed. The growth of the azimuthal anchoring energy was observed for all parameters (laser power $P$ and scanning speed $\upsilon$) of the NLL method. However, the gain effect to the value $\sim 10^{-4}$ J/m$^2$ (Fig. 11 (a), (b)) can be reached for the scanning speed $\upsilon = 1500$ mm/s and laser power $P$ changing within a range 300 – 350 mW. From Figure 11 it may be concluded that the choice of both the scanning speed and laser power and additional using of polymer-coated films make it possible to change the value of the azimuthal anchoring energy within a wide range $\sim 10^{-6} - 10^{-4}$ J/m$^2$.

### 4 Conclusions

In this manuscript, for the first time, we studied in detail the aligning surfaces obtained by the nonlinear laser lithography (NLL), which was recently proposed in [41], as a new alternative technique for the alignment of nematic LCs. This technique for the LC alignment is simple, high-speed and low-cost to create large areas of various surfaces with microgrooves. In our experiments, the measured depth of grooves can be changed by a laser power. The azimuthal anchoring energy of the NLL-induced microgrooves of the Ti layer is of order of photoaligning layers. It was experimentally shown that the azimuthal anchoring energy depends on both the scanning speed and laser power, which are main controlled parameters of the NLL technique. We show the possibility of a gain effect of the anchoring energy of the microstructured Ti layers, owing to the coating of a polymer ODAPI film onto structured surfaces. The composite-aligning surface, based on the microstructured Ti layer coated with the ODAPI film, is an analogue to the surface with a rubbed polymer. However, the creation of an aligning layer by the NLL method has certain difference as to the rubbing or photoaligning technique. Namely, at the first stage, a grooves structure is created. At the second stage, the magnification of the azimuthal anchoring energy occurs, by coating with a polymer of the grooves structure followed by the polymerization process. It was shown that the microstructured Ti layer is characterized with a relatively weak azimuthal anchoring energy, while by coating with a polymer, a Ti layer has a strong anchoring energy. It was experimentally shown that the value of the azimuthal anchoring energy in a wide range can be controlled by means of changing at least two NLL parameters (scanning speed and laser power) during the structuring of the Ti layer and further coating with a polymer film without additional processing. This modified method, based on the processing of metal layers by application of the NLL method, provide an alternative comparable to the existing techniques, such as rubbing and photoaligning methods.

**Acknowledgments**

The authors thank W. Becker (Merck, Darmstadt, Germany) for his generous gift of the nematic liquid crystal E7, Prof. I. Gerus (Institute of Bio-organic Chemistry and Petrochemistry, NAS of Ukraine) for the kind provision of the ODAPI polymer, and Dr. P. Lytvyn and Dr. A. Korchovyi (V. E. Lashkaryov Institute of Semiconductor Physics, NAS of Ukraine) for their help during AFM studies. I.P. thanks M. Gure and E. Karaman (Bilkent University, Turkey) for the technical support of the NLL method. I. P. and F. Ö. I. acknowledge financial support from FP7 European Research Council (ERC) (NLL - 617521).

**References**

[1]  J. Congard, Alignment of Nematic Liquid Crystals and Their Mixtures, Gordon and Breach Science Publishers, London, New York, Paris, 1982.

[2]  K. Ishimura, Photoalignment of liquid-crystals systems, Chem. Rev. 100 (2000) 1847-1873. Doi: 10.1021/cr980079e

[3]  M.O'Neill, S.M. Kelly, Photoinduced surface alignment for liquid crystal displays, J. Phys. D: Appl. Phys.. 33 (2000) R67-R84. Doi: 10.1088/0022-3727/33/10/201

[4]  O. Yaroshchuk, Yu. Reznikov, Photoalignment of liquid crystals: basic and current trends, J. Mater. Chem. 22 (2012) 286-300. Doi: 10.1039/C1JM13485J

[5]  T. Seki, New strategies and implications for the photoalignment of liquid crystalline polymers, Polymer Journal. 46 (2014) 751-768. Doi: 10.1038/pj.2014.68

[5]  D.W. Berreman, Solid surface shape and the alignment of an adjacent nematic liquid crystal, Phys. Rev. Lett. 28 (1972) 1683-1686. Doi: 10.1103/PhysRevLett.28.1683

[6]  D.W. Berreman, Alignment of liquid crystals by grooves surfaces, Mol. Cryst. Liq. Cryst. 23 (1973) 215-231. Doi: 10.1080/15421407308083374.

[7]  P. Chatelain, Orientation of liquid crystals by rubbed surfaces, C. R. Acad. Sci. 231 (1941) 875-876

[8]  T.J. Sluckin, D.A. Dunmur, H. Stegemeyer, eds., Crystal That Flow. Classic Papers from History of Liquid Crystals, CRC Press, Taylor & Francis Group, 2004. https://doi.org/10.1201/9780203022658

[9]  A. Yasutake, R.K. Oshiro, Dispositif d'affichage à cristaux liquids et son procédé de réalization, France patent application, N7536984 (1975).

[10] K. Ishimura, Y. Suzuki, T. Seki, A. Hosori, K. Aoki, Reversible change in alignment mode of nematic liquid crystals regulated photochemically by command surfaces modified with an azobenzene monolayer, Langmuir. 4 (1988) 1214-1216. Doi: 10.1021/la00083a030

[11] W. M. Gibbons, P. J. Shannon, S. T. Sun, B. J. Swetlin, Surface-mediated alignment of nematic liquid crystals with polarized laser light, Nature. 351 (1991) 49-50. Doi: 10.1038/351049a0

[12] M. Schadt, K. Shmitt, V. Kozinkov, V. Chigrinov, Surface-induced parallel alignment of liquid crystals by linearly polymerized photopolymers, Jpn. J. Appl. Phys. 31 (1992) 2155-2164. Doi: 10.1143/JJAP.31.2155

[13] A. Dyadyusha, V. Kozinkov, T. Marusii, Y. Reznikov, V. Reshetnyak, A. Khizhnyak, Light-induced planar alignment of nematic liquid-crystal by the anisotropic surface without mechanical texture, Ukr. Fiz. Zh. 36 (1991) 1059-1062.

[14] A.G. Dyadyusha, T.Ya. Marusii, V.Yu. Reshetnyak, Yu.A. Reznikov, A.I. Khizhnyak, Orientational effect due to a change in the anisotropy of the interaction between a liquid crystal and a bounding surface, JETF Lett. 56 (1992) 17-21.

[15] D. Voloshchenko, A. Khizhnyak, Yu. Reznikov, V. Reshetnyak, Control of an easy-axis on nematic-polymer interface by light action to nematic bulk, Jpn. J. Appl. Phys. 34 (1995) 566-571. Doi: 10.1143/JJAP.34.566

[16] O. Yaroshchuk, R. Kravchuk, A. Dobrovolskyy, L. Qiu, O. Lavrentovich, Planar and tilted uniform alignment of liquid crystals by plasma-treated substrates, Liquid Crystals. 31 (2004) 859-869. Doi: 10.1080/02678290410001703145

[17] O. Yaroshchuk, Yu. Zakrevskyy, A. Dobrovolskyy, S. Pavlov, Liquid crystal alignment on the polymer substrates irradiated by plasma beam, Proc. SPIE, 4418 (2001) 49-53.

[18] C.-R. Lee, T.-L.Fu, K.-T. Cheng, T.-S. Mo, A. Y.-G. Fuh, Surface-assisted photoalignment in due-doped liquid-crystal films, Phys. Rev. E. 69 (2004) 031704-1-6. Doi: 10.1103/PhysRevE.69.031704

[19] A.Y.-G. Fuh, C.-K. Liu, K.-T. Cheng, C.-L. Ting, C.-C. Chen, P.C.-P. Chao, H.-K. Hsu, Variable liquid crystal pretilt angles generated by photoalignment in homeotropically aligned azo dye-doped liquid crystals, Appl. Phys. Lett. 95 (2009) 161104-1-3. Doi: 10.1063/1.3253413

[20] D.M. Tennant, T.L. Koch, P.P. Mulgrew, R.P. Gnall, F. Ostermeyer, J.-M. Verdiell, Characterization of near-field holography grating mask for optoelectronics fabricated by electron beam lithography, J. Vac. Sci. Technol. B. 10 (1992) 2530-2535. Doi: 10.1116/1.586052

[21] J.-H. Kim, M. Yoneya, H. Yakoyama, Tristable nematic liquid-crystal device using micropatterned surface alignment, Nature. 420 (2002) 159-162. Doi: 10.1038/nature01163

[22] S.Y. Chou, P.R. Krauss, P.J. Renstrom, Imprint lithography with 25-nanometer resolution. Science. 272 (1996) 85-87. DOI: 10.1126/science.272.5258.85

[23] K.O. Hill, B. Malo, F. Bilodeau, D.C. Johnson, J. Albert, Bragg gratings fabricated in monomode photosensitive optical fiber by UV exposure through a phase mask, Appl. Phys. Lett. 62 (1993) 1035-1037. Doi: 10.1063/1.108786

[24] T.-C. Lin, L.-C. Huang, T.-R. Chou, C.-Y. Chao, Alignment control of liquid crystal molecules using crack induced self-assembled grooves, Soft Matt. 5 (2009) 3672-3676. Doi: 10.1039/B911567F

[25] B. Öktem, I. Pavlov, S. Ilday, H. Kalaycıoğlu, A. Rybak, S. Yavas, M. Erdoğan, F.Ö. Ilday, Nonlinear laser lithography for indefinitely large-area nanostructuring with femtosecond pulse, Nat. Photonics. 7 (2013) 897-901. Doi: 10.1038/NPHOTON.2013.272

[26] Licristal brochure, Merck Liquid crystals, 1994.

[27] E.P. Rynes, C.V. Brown, J.F. Strömer, Method for the measurement of the $K_{22}$ nematic elastic constant, App Phys Lett. 82 (2003) 13–15. Doi: 10.1063/1.1534942

[28] F. Yang, J.R. Sambles, G.W. Bradberry, Half-leaky guided wave determination of azhimuthal anchoring energy and twist elastic constant of a homogeneously aligned nematic liquid crystal, J Appl Phys. 85 (1999) 728–733. Doi: 10.1063/1.369153

[29] I. Pavlov, A. Rybak, Ç. Şenel, F. Ö. Ilday, Balancing gain narrowing with self phase modulation: 100-fs, 800-nJ from an all-fiber-integrated Yb amplifier, The European Conference on Lasers and Electro-Optics, (2013) CJ_6_5.

[30]   J. Bonse, J. Krüger, S. Höhm, A. Rosenfeld, Femtosecond laser-induced periodic surface structures, Journal of Laser Applications. 24 (2012) 042006-1-7. Doi: 10.2351/1.4712658

[31]   J. Bonse, S. Höhm, S. V. Kirner, A. Rosenfeld, J. Krüger, Laser-induced periodic surface structures - a scientific evergreen, IEEE Journal of Selected Topics in Quantum Electronics. 23 (2017) 1-15. Doi: 10.1109/JSTQE.2016.2614183

[32]   D. Andrienko, Yu. Kurioz, M. Nishikawa, Yu. Reznikov, J.L. West, Control of the anchoring energy of rubbed polyimide layers by irradiation with depolarized UV light, Jpn. J. Appl. Phys. 39 (2000) 1217-1220. Doi: 10.1143/JJAP.39.1217

[33]   I. Gerus, A. Glushchenko, S.-B. Kwon, V. Reshetnyak, Yu. Reznikov, Anchoring of a liquid crystal on photoaligning layer with varying surface morphology, Liquid Crystals. 28 (2001) 1709-1713. Doi: 10.1080/02678290110076371

[34]   I. Gvozdovskyy, Electro- and photoswitching of undulation structures in planar cholesteric layers aligned by a polyimide film possessing various values of anchoring energy, Liquid Crystals. xx (2017) 1-17 (In press). Doi:10.1080/02678292.2017.1359691.

[35]   B.I. Senyuk, I.I. Smalyukh, O.D. Lavrentovich, Undulations of lamellar liquid crystals in cells with finite surface anchoring near and well above the threshold, Phys. Rev. E. 74 (2006) 011712-1-13. Doi: 10.1103/PhysRevE.74.011712

[36]   I.I. Smalyukh, O.D. Lavrentovich, Anchoring-mediated interaction of edge dislocations with bounding surfaces in confined cholesteric liquid crystals, Phys. Rev. Lett. 90 (2003) 085503-1-4. Doi: 10.1103/PhysRevLett.90.085503

[37]   T.K. Gaylord, M.G. Moharam, Thin and thick gratings: terminology clarification, Appl Opt. 20 (1981) 3271-3273. Doi: 10.1364/AO.20.003271

[38]   V.G. Chigrinov, V.M. Kozenkov, H.S. Kwok, eds., Photoalignment of Liquid Crystalline Materials: Physics and Applications, A John Wiley & Sons, Ltd., Publication, England, 2008.

[39]   Yu.V. Kozenkov, V.G. Chigrinov, S.-B. Kwon, Photoanisotropic effects in poly(vinyl-cinnamate) derivatives and their applications, Mol. Cryst. Liq. Cryst. 409 (2004) 251-267. Doi: 10.1080/15421400490431900

[40]   G.P. Bryan-brown, I.C. Sage, Photoinduced ordering and alignment properties of polyvinylcinnamates, Liquid Crystals. 20 (1996) 825-829. Doi: 10.1080/02678299608033178

[41]   I. Pavlov, A. Rybak, A. Dobrovolskiy, V. Kadan, I. Blonskiy, F. Ö. Ilday, Z. Kazantseva, I. Gvozdovskyy, The alignment of nematic liquid crystal by the Ti layer processed by nonlinear laser lithography, arXiv:1707.03612v2.

**Figure captions**

Fig. 1. (a) Scheme of the structuring of the Ti layer by NLL method. (b) Cartoon, demonstrating scanning direction of the laser beam over the sample during NLL process. (c) Photograph of the sample with ten structured areas, and SEM image of the square area with dimension 5×5 mm$^2$.

Fig. 2. Schematic image of the combined twist LC cell, consisting of the reference substrate (rubbed PI2555 film) and tested substrate (microstructured pure Ti layer or coated with ODAPI film). Direction of the microgrooves structured Ti layer $\theta_1$ is inclined at 8º to the horizontal side of the square area. Direction of rubbing of the reference substrate $\theta_2$ is at 45º with the horizontal side of the square area (microstructured Ti layer) of the tested plate.

Fig. 3. Scheme of the measurement of the twist angle $\varphi$ of the combined twist LC cell, assembled with the reference PI2555 ($N_{rubb}$ = 10) and tested substrates of both types.

Fig. 4. (a) AFM image of the structured Ti layer after processing by the NLL method with a speed of scanning $\upsilon$ = 1500 mm/s and laser power $P$ = 350 mW. (b) Dependence of the depth of grooves A on laser power $P$ at a constant scanning speed $\upsilon$ = 1500 mm/s. (c) Dependence of the period of grooves $\Lambda$ on a scanning speed at a constant laser power $P$ = 350 mW, measured by the AFM method (solid circles) and diffraction method (open squares). The dashed line is a guide to the eye.

Fig. 5. Dependence of the anchoring energy, calculated by Berreman's theory, of the aligning layer on: (a) period (open symbols) and (b) depth of grooves (solid symbols). The dashed line is a guide to the eye.

Fig. 6. Dependencies of the anchoring energy $W_B$ of the microstructured Ti layer on: (a) the power of laser beam $P$ and (b) depth of grooves A. The period of grooves of the microstructured Ti layer is 920 nm, and the scanning speed is 1500 mm/s. The dashed line is a guide to the eye.

Fig. 7. Photographs of the twist LC cells placed between a pair of polarizers with different angles of planes of polarization: (a), (d) - parallel; (b), (e) – twist angle $\varphi'$; (c), (f) - perpendicular. The LC cells consist of the reference (rubbed PI2555 film) and tested substrate, having the microstructured Ti layer (a-d) and microstructured Ti layer coated with the ODAPI film (d-f). The twist angle $\varphi'$ for the LC cell: (a-c) - 22º; (d-f) – 33.5º. The thickness of the twisted LC cell was: (a) – (c) $d$ = 23.3 μm; (d) – (f) $d$ = 23.8 μm. The processing of Ti layers was carried out by the NLL method at the scanning speed $\upsilon$ = 500 mm/s and laser power $P$ = 250 mW.

Fig. 8. Photographs of the twist LC cells placed between parallel polarizers. LC cells consist with a pair of substrates: (a) the reference glass plate is the rubbed PI2555 film and tested plate is the non-rubbed ODAPI film formed by the dipping technique; (b) the reference plate is the non-rubbed PI2555 film and tested plate is the microstructured Ti layer coated with the ODAPI film by means of the dipping technique. Thickness of LC cells was ~ 24.2 μm.

Fig. 9. Dependence of the twist angle $\varphi$ of the LC cell on: (a) the scanning speed $v$ and (b) laser power $P$. The LC cell consists of the both types of the tested substrate: microstructured Ti layer (solid squares and open diamonds) and microstructured Ti layer coated with the ODAPI film (solid circles and open triangles). The dashed line is a guide to the eye.

Fig. 10. Dependence of the calculated azimuthal anchoring energy $W_\varphi$ on the scanning speed $v$ for LC cells consisting of the both types of the tested substrates, having the microstructured Ti layer (solid triangles) and microstructured Ti layer coated with the ODAPI film (solid spheres). The inset depicts the dependence of $W_\varphi(v)$ for the microstructured Ti layer in a large scale. The dashed line is a guide to the eye.

Fig. 11. Dependence of the calculated azimuthal AE (Eq.1) of aligning films on the power of the laser beam $P$ at a constant scanning speed $v$: (a) 500 mm/s, (b) 1500 mm/s and (c) 2300 mm/s during the processing of the Ti layer with the NLL method. The LC cell consists of the tested substrate having the microstructured Ti layer (solid triangles) and structured Ti layer coated with the ODAPI film (solid spheres). The dashed line is a guide to the eye.